\documentclass[aps,prl,preprint,nopacs,superscriptaddress]{revtex4}
\usepackage{amsmath}
\usepackage{amssymb}
\usepackage{graphicx}
\usepackage{hyperref}
\usepackage[utf8]{inputenc}
\newcommand{\beq}{\begin{equation*}}
\newcommand{\eeq}{\end{equation*}}
\newcommand{\s}{Co$_3$Sn$_2$S$_2$}
\newcommand{\f}{$E_\textrm{F}$}

\newcommand{\ang}{${\rm \AA}$}

\newcommand{\ab}{{\it ab initio}}
\newcommand{\Ab}{{\it Ab initio}}

\newcommand{\pana}{(a)}
\newcommand{\panb}{(b)}
\newcommand{\panc}{(c)}
\newcommand{\pand}{(d)}

\newcommand{\panab}{(a,b)}

\newcommand{\cpana}{({\bf a})}
\newcommand{\cpanb}{({\bf b})}
\newcommand{\cpanc}{({\bf c})}
\newcommand{\cpand}{({\bf d})}

\begin{document}

\title{Signatures of a topological Weyl loop in \s}

\author{Ilya Belopolski\footnote{These authors contributed equally to this work.}} \email{ilyab@princeton.edu}
\affiliation{Laboratory for Topological Quantum Matter and Spectroscopy (B7), Department of Physics, Princeton University, Princeton, New Jersey 08544, USA}

\author{Tyler A. Cochran$^*$}
\affiliation{Laboratory for Topological Quantum Matter and Spectroscopy (B7), Department of Physics, Princeton University, Princeton, New Jersey 08544, USA}

\author{Stepan S. Tsirkin}
\affiliation{Department of Physics, University of Zurich, Winterthurerstrasse 190, 8057 Zurich, Switzerland}

\author{Zurab Guguchia}
\affiliation{Laboratory for Topological Quantum Matter and Spectroscopy (B7), Department of Physics, Princeton University, Princeton, New Jersey 08544, USA}
\affiliation{Laboratory for Muon Spin Spectroscopy, Paul Scherrer Institute, Villigen PSI, Switzerland}

\author{Jiaxin Yin}
\affiliation{Laboratory for Topological Quantum Matter and Spectroscopy (B7), Department of Physics, Princeton University, Princeton, New Jersey 08544, USA}

\author{Songtian S. Zhang}
\affiliation{Laboratory for Topological Quantum Matter and Spectroscopy (B7), Department of Physics, Princeton University, Princeton, New Jersey 08544, USA}

\author{Z\v{\i}j\={\i}a Ch\'eng}
\affiliation{Laboratory for Topological Quantum Matter and Spectroscopy (B7), Department of Physics, Princeton University, Princeton, New Jersey 08544, USA}

\author{Xiaoxiong Liu}
\affiliation{Department of Physics, University of Zurich, Winterthurerstrasse 190, 8057 Zurich, Switzerland}

\author{Guoqing Chang}
\affiliation{Laboratory for Topological Quantum Matter and Spectroscopy (B7), Department of Physics, Princeton University, Princeton, New Jersey 08544, USA}

\author{Xi\`an Y\'ang}
\affiliation{Laboratory for Topological Quantum Matter and Spectroscopy (B7), Department of Physics, Princeton University, Princeton, New Jersey 08544, USA}

\author{Daniel Multer}
\affiliation{Laboratory for Topological Quantum Matter and Spectroscopy (B7), Department of Physics, Princeton University, Princeton, New Jersey 08544, USA}

\author{Timur K. Kim}
\affiliation{Diamond Light Source, Didcot OX11 0DE, UK}

\author{Cephise Cacho}
\affiliation{Diamond Light Source, Didcot OX11 0DE, UK}

\author{Claudia Felser}
\affiliation{Max Planck Institute for Chemical Physics of Solids, N\"othnitzer Stra{\ss}e 40, 01187 Dresden, Germany}

\author{Titus Neupert}
\affiliation{Department of Physics, University of Zurich, Winterthurerstrasse 190, 8057 Zurich, Switzerland}

\author{M. Zahid Hasan} \email{mzhasan@princeton.edu}
\affiliation{Laboratory for Topological Quantum Matter and Spectroscopy (B7), Department of Physics, Princeton University, Princeton, New Jersey 08544, USA}
\affiliation{Princeton Institute for Science and Technology of Materials, Princeton University, Princeton, New Jersey, 08544, USA}
\affiliation{Materials Sciences Division, Lawrence Berkeley National Laboratory, Berkeley, CA 94720, USA}


\begin{abstract}
The search for novel topological phases of matter in quantum magnets has emerged as a frontier of condensed matter physics. Here we use state-of-the-art angle-resolved photoemission spectroscopy (ARPES) to investigate single crystals of \s\ in its ferromagnetic phase. We report for the first time signatures of a topological Weyl loop. From fundamental symmetry considerations, this magnetic Weyl loop is expected to be gapless if spin-orbit coupling (SOC) is strictly zero but gapped, with possible Weyl points, under finite SOC. We point out that high-resolution ARPES results to date cannot unambiguously resolve the SOC gap anywhere along the Weyl loop, leaving open the possibility that \s\ hosts zero Weyl points or some non-zero number of Weyl points. On the surface of our samples, we further observe a possible Fermi arc, but we are unable to clearly verify its topological nature using the established counting criteria. As a result, we argue that from the point of view of photoemission spectroscopy the presence of Weyl points and Fermi arcs in \s\ remains ambiguous. Our results have implications for ongoing investigations of \s\ and other topological magnets.
\end{abstract}

\date{\today}
\maketitle

\begin{figure}
\centering
\includegraphics[width=10cm,trim={0in 0in 0in 0in},clip]{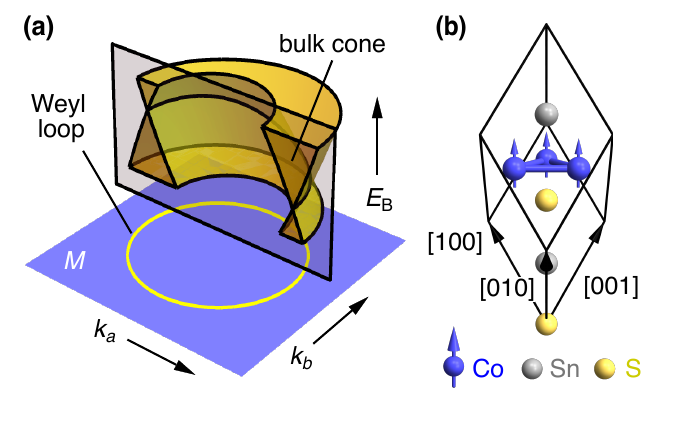}
\caption{\label{intro} {\bf Weyl loops, ferromagnetism \& mirror symmetry.} \cpana\ Schematic of a Weyl loop electronic structure \cite{Co2MnGa_me}, contained in a mirror plane $M$ (purple) in the bulk three-dimensional Brillouin zone. \cpanb\ Primitive unit cell of \s, consisting of one formula unit (7 atoms), displayed so that the crystallographic mirror plane is apparent (perpendicular to the page, cutting down the center of the cell). The conventional unit cell $z$ direction (layer stacking direction) is $[111]$ in the primitive basis.}
\end{figure}

Quantum magnets with electronic topology have emerged as a promising arena for novel topological states, magnetic control of topological invariants and anomalous transport \cite{news_daSilvaNeto,ReviewMagTopo_Tokura,RMPWeylDirac_Armitage,ReviewQuantumMaterials_Hsieh,ReviewQuantumMaterials_Nagaosa,ARCMP_me,ARCMP_Xue,Fe3Sn2_Checkelsky,Fe3Sn2_Jiaxin,Co2MnGa_me}. One important magnetic topological structure is the Weyl loop (or Weyl line), where two singly-degenerate bands cross along a closed loop in momentum space, Fig. \ref{intro}\pana\  \cite{Co2MnGa_me,WeylLoopSuperconductor_Nandkishore,WeylLines_Kane,WeylDiracLoop_Nandkishore}. Weyl loops arise naturally under the combination of ferromagnetism and crystalline mirror symmetry. Ferromagnetic order is associated with a spin splitting, so that bands are generically singly-degenerate throughout the Brillouin zone. At the same time, mirror symmetry can protect two-fold degenerate band crossings taking the form of closed loops living in the Brillouin zone mirror planes. The result is a bulk cone dispersion which persists along a closed loop in momentum space. Since ferromagnetism is common and many crystal structures have mirror symmetry, Weyl loops are expected to arise naturally in many quantum magnets. Such Weyl loops are typically expected to be gapless only when spin-orbit coupling is strictly zero. With non-zero spin-orbit coupling, the Weyl loop is expected to gap out, although it can leave behind some gapless Weyl points. Weyl loops are associated with singularities of Berry curvature, which can give rise to a large anomalous Hall effect and other exotic transport response, even when the singularity is made well-defined by spin-orbit coupling \cite{Co2MnGa_me}.

The kagome magnet \s\ has recently captured the attention of the community, particularly with the observation of a large anomalous Hall effect \cite{Co3Sn2S2_HechangLei,Co3Sn2S2_Enke}. \s\ crystallizes in space group $R\bar{3}2/m$ (No. 166), with dihedral point group $D_{3d}$ which includes three mirror planes, Fig. \ref{intro}\panb. Neutron scattering and muon spin rotation experiments at low temperature suggest a ferromagnetic order with magnetic moments associated with the Co atoms and oriented along the $z$ direction, $[111]$ in the primitive basis \cite{Co3Sn2S2_Zurab,Co3Sn2S2_Sobany,Co3Sn2S2_Weihrich}. \Ab\ calculations under ferromagnetic order predicted that \s\ exhibits Weyl loops living in the mirror planes, neglecting spin-orbit coupling  \cite{Co3Sn2S2_QiunanXu, Co3Sn2S2_HechangLei}. Under spin-orbit coupling, these Weyl loops are predicted to gap out almost everywhere, leaving behind several Weyl points. Recent studies have explored various exotic properties of \s, including offering experimental evidence for Weyl points and topological Fermi arcs by angle-resolved photoemission spectroscopy (ARPES) \cite{Co3Sn2S2_HechangLei, Co3Sn2S2_Jiaxin, Co3Sn2S2_YulinChen, Co3Sn2S2_Beidenkopf, Co3Sn2S2_Zurab, Co3Sn2S2_me_APS2019}. However, the existing ARPES results have been limited by broad spectral linewidth \cite{Co3Sn2S2_HechangLei}; the absence of photon energy dependence on bulk electronic states \cite{Co3Sn2S2_YulinChen, Co3Sn2S2_Zurab}; lack of a surface state Chern number counting \cite{Co3Sn2S2_HechangLei, Co3Sn2S2_YulinChen}; and lack of an investigation of the spin-orbit coupling gap in the Weyl loop, which is crucial for identifying possible Weyl points \cite{Co3Sn2S2_HechangLei, Co3Sn2S2_YulinChen, Co3Sn2S2_Zurab}. These missing pieces of the puzzle motivate further investigation of the topology of the ferromagnetic electronic structure in \s.

We grew high-quality single crystals of \s\ by a self-flux method with congruent composition \cite{Co3Sn2S2_Enke}. We carried out photon-energy-dependent ARPES measurements at Beamline I05 of Diamond Light Source, Harwell Science Campus, Oxfordshire, UK using a Scienta R4000 electron analyzer with angular resolution $< 0.2^{\circ}$; total energy resolution $< 13$ meV for all photon energies; spot size 50 $\mu$m $\times$ 50 $\mu$m; and sample temperature 8 K. \cite{Diamond_Hoesch}. We carried out additional ARPES measurements at Beamline 5-2 of the Stanford Synchrotron Radiation Lightsource, SLAC in Menlo Park, CA, USA using a Scienta R4000 electron analyzer; photon energy $h\nu = 130$ eV; angular resolution $ < 0.2^{\circ}$; beam spot size 16 $\mu$m (vertical) $\times$ 36 $\mu$m (horizontal); and sample temperature 20 K. Samples were cleaved $\textit{in situ}$ and measured under a vacuum of $4 \times 10^{-11}$ Torr or better. We studied the $[111]$ (primitive lattice basis), $z$ direction, surface of the crystal. Data were symmetrized and a background subtraction was carried out. Density functional theory (DFT) calculations with the projected augmented wave (PAW) method were implemented in the Vienna \ab\ simulation package (VASP) \cite{DFT2,DFT_Efficiency_KressFurthmueller} with generalized gradient approximation (GGA) \cite{DFT4}. The surface spectral function was calculated for the Sn-terminated surface from a tight-binding model derived from maximally localized Wannier functions \cite{Yates_Wannier90}, using the WannierTools package \cite{Soluyanov_WannierTools}.

\begin{figure}
\centering
\includegraphics[width=13cm,trim={0in 0.3in 0in 0.2in},clip]{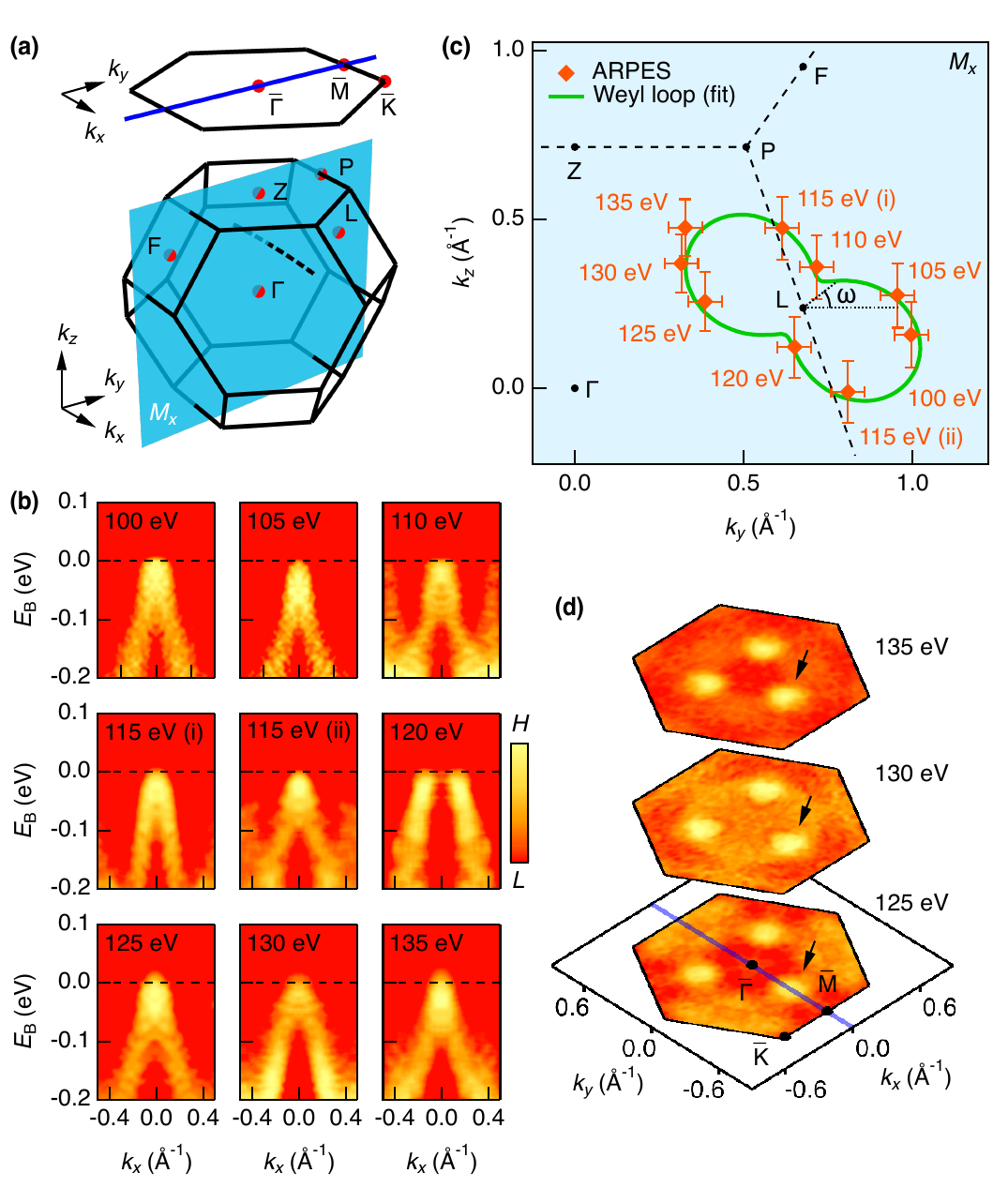}
\caption{\label{line} {\bf Weyl loop in \s.} \cpana\ Bulk and $(111)$ surface Brillouin zones of \s\ with bulk mirror plane ($k_y$, $k_z$ plane, cyan), surface mirror line (blue) and several high-symmetry points (red) marked. The surface zone corresponds to the natural cleaving plane. \cpanb\ Energy-momentum ARPES spectra acquired perpendicular to the mirror plane at different photon energies $h\nu$, with $k_y$ locations as marked in \panc. Linear vertical light polarization, $\bar{\Gamma}-\bar{K}$ ARPES slit alignment, first in-plane Brillouin zone. \cpanc\ Bulk momentum-space locations of cone crossings extracted from the spectra (orange dots), along with a fit to the orange dots using a polar coordinate Fourier decomposition around the $L$ point (green curve, see main text). The ARPES out-of-plane momentum $k_z$ corresponds approximately to $h\nu$; $k_y$ is an ARPES in-plane momentum. The plane of \panc\ is the $M_x$ mirror plane. The polar angle $\omega$ takes $L$ as the origin. \cpand\ Isoenergy surfaces at the Fermi level, acquired by ARPES at several $h\nu$, showing a dot-like Fermi surface (black arrows).}
\end{figure}

We explore the ARPES spectra of our \s\ samples along momentum slices perpendicular to the Brillouin zone mirror plane, see dotted line in Fig. \ref{line}\pana. Measuring with incident photon energy $h\nu = 130$ eV, we observe a cone-like dispersion near the Fermi level \f\ at $k_y = 0.32\ \ang^{-1}$, with the top of the cone near $k_x = 0\ \ang^{-1}$, Fig. \ref{line}\panb. Examining next the ARPES isoenergy contours, we observe a point-like electronic structure at the Fermi level, suggestive of a band crossing occurring on the mirror plane $M_x$, Fig. \ref{line}\pand. To better understand this cone feature, we repeat our measurement at $h\nu = 125$ eV. We again observe a cone-like dispersion centered on $M_x$, with again a point-like Fermi surface, Figs. \ref{line}\panb, \pand, but with $k_y$ shifted to $0.39\ \ang^{-1}$. Our observation that the cone persists with varying photon energy, although the cone does shift in $k_y$, points to a three-dimensional nodal electronic state extending along the out-of-plane $k_z$ momentum direction. To better understand these results, we systematically track the cone dispersion in our ARPES spectra from $h\nu = 100$ to $135$ eV, Fig. \ref{line}\panb. We again find that the cone persists with photon energy, with the crossing point consistently on the $M_x$ plane, but at varying $k_y$. To summarize this photon energy dependence, we assemble the crossing points of all cones in the $(k_y, k_z)$ mirror plane, where larger $k_z$ are obtained by measurements with higher photon energy, Fig. \ref{line}\panc. We observe that the crossing points appear to encircle the L point of the bulk Brillouin zone, suggesting that we have observed a bulk line node on $M_x$ in \s. Since the system is ferromagnetic with generically singly-degenerate bands, we interpret this line node as a Weyl loop, Fig. \ref{intro}\pana. To extract the complete trajectory of the Weyl loop, we parametrize its trajectory by an angle $\omega$ in polar coordinates with L taken as the origin. The Weyl loop is then described by a function $r (\omega)$ with $2\pi$ periodicity. Crystalline inversion symmetry $P$ further requires that the dispersion remain unchanged under inversion through $L$, constraining the trajectory to $r (\omega + \pi) = r (\omega)$. With $\pi$ periodicity, the first two terms of the Fourier decomposition are $r (\omega) = r_0 + r_1 \cos (2\omega + \phi_1)$. By fitting to the ARPES locations of the cone dispersions, we find that the trajectory of the Weyl loop is given by $r_0 = 0.24\ {\rm \AA}^{-1}$, $r_1 = 0.14\ {\rm \AA}^{-1}$ and $\phi_1 = 63^{\circ}$, Fig. \ref{line}\panc. Our analysis provides the first example of the full momentum-space trajectory of a Weyl loop, extracted from experimental data alone.

Having provided evidence for a topological Weyl loop in the ferromagnetic phase of \s, we next consider the behavior of the Weyl loop under spin-orbit coupling. Theoretically, with non-zero spin-orbit coupling, allowed perturbations can lift the two-fold degeneracy and open a gap in the Weyl loop. Depending on details of the electronic structure, qualitatively different behaviors are possible: either the Weyl loop may gap out fully along its entire trajectory; or some points may remain gapless, forming a number of discrete Weyl points. To distinguish between these scenarios, it is necessary to investigate the spin-orbit gap along the Weyl loop. Looking again at our ARPES spectra, we find that at certain photon energies, such as $h\nu = 120$ eV, the Fermi level lies well below the crossing point, preventing us from investigating the spin-orbit gap, Fig. \ref{line}\panb. At other photon energies, such as $h\nu = 130$ eV, the crossing point appears to be close to the Fermi level, but still no upper Weyl cone is visible and spectral weight persists to the Fermi level, again obscuring any possible spin-orbit gap. In fact, we observe no clear spin-orbit gap anywhere along the Weyl loop trajectory. Therefore, although the Weyl loop is apparent in our photoemission experiments, we cannot clearly demonstrate the presence or absence of Weyl points in \s.

\begin{figure}
\centering
\includegraphics[width=17cm,trim={0in 0in 0in 0in},clip]{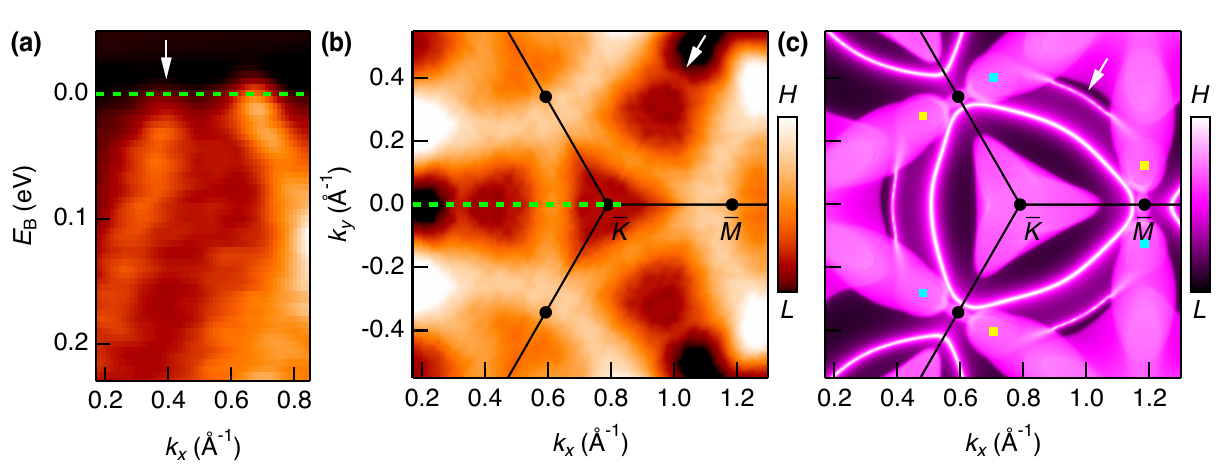}
\caption{\label{arc} {\bf Surface state analysis in \s.} \cpana\ Energy-momentum ARPES spectrum along $\bar{\Gamma}-\bar{K}$ and \cpanb\ ARPES Fermi surface in the vicinity of the $\bar{K}$ point, acquired at $h\nu = 130$ eV, linear horizontal light polarization, $\bar{\Gamma}-\bar{M}$ APRES slit alignment, second/third surface Brillouin zones. White arrows indicate a candidate topological Fermi arc. \cpanc\ \Ab\ calculation of the surface density of states at the Fermi energy, Sn termination, with spin-orbit coupling included. Weyl point predictions marked by the yellow (positive chirality) and cyan (negative chirality) squares.}
\end{figure}

Lastly, we consider the topological Fermi arc surface states as a separate avenue by which to demonstrate a Weyl point semimetal phase in \s. We find that it remains difficult to pinpoint topological Fermi arcs in \s. In our Fermi surface acquired by photoemission spectroscopy, we observe (1) a rather sharp triangular state around the $\bar{K}$ point; (2) a second set of faint, but rather sharp, arc-like states closer to $\bar{\Gamma}$; and (3) several wide regions of broad spectral weight, for example around $\bar{M}$, Figs. \ref{arc}\panab. We compare these spectra with an \ab\ calculation of the (111) surface of \s, Sn termination, taking into account spin-orbit coupling. In the calculation, we observe a qualitative match to our experimental results, including (1) a trivial surface state encircling the $\bar{K}$ point; (2) a nascent topological Fermi arc connecting the calculated Weyl cone projections; and (3) several projected bulk pockets, which include bulk states associated with the predicted Weyl points, see the yellow and cyan squares, Fig. \ref{arc}\panc. To demonstrate the presence or absence of a topological Fermi arc, we attempt to apply the surface state analysis criteria directly to our photoemission spectra, focusing in particular on the arc-like state \cite{NbP_me}. None of the observed states, including the arc-like state, take the form of separated, disjoint arcs (Criterion 1, \cite{NbP_me}) nor do we observe any apparent kinks (Criterion 2) or an odd set of closed contours (Criterion 3). Next we consider how one might build a momentum-space Chern counting path enclosing a possible Weyl point (Criterion 4). In our photoemission spectrum along the $\bar{\Gamma}-\bar{K}$ path, the arc-like state exhibits a clear right-moving Fermi velocity, contributing a count of $+1$, Fig. \ref{arc}\pana. To close this momentum-space path, we need to find another way back around to the starting point without entering any bulk projection. Our experimental data suggests that there is no momentum-space path available which encloses a candidate Weyl pocket and could signal a non-zero Chern number. All relevant paths would cross some bulk states, so we could not rule out additional boundary modes above the Fermi level which change the net Chern number count. \Ab\ results also suggest that there is no suitable momentum-space path available at or below the Fermi level for Chern number counting, Fig. \ref{arc}\panc. At the same time, our \ab\ calculations do suggest that a `persistent' topological Fermi arc survives in \s, despite being enclosed by and partially degenerate with bulk projection. Such a `persistent' topological surface state has also been experimentally observed at least once, in another system \cite{HgTe_ChangLiu}. An analogous situation may arise in \s, but further careful study is needed. Given the current theoretical and experimental limitations, it remains unclear whether or not \s\ exhibits topological Fermi arcs.

We have asked what can be reasonably concluded about the topology of \s\ based only on our ARPES spectra and basic symmetry considerations, as opposed to an {\it ab-initio-}led approach. The takeaway is that the ARPES spectra suggest the presence of a Weyl loop in \s\ in the ferromagnetic phase. However, it remains unclear if the Weyl loop under spin-orbit coupling leaves behind Weyl points or is fully gapped throughout. Similarly, an ARPES surface state analysis fails to reveal unambiguous topological Fermi arcs. Nonetheless, our ARPES spectra show that the Weyl loop lies near the Fermi level along most of its momentum-space trajectory, suggesting that the large anomalous Hall effect in \s\ may be understood as arising from a topological Weyl loop. This interpretation appears to be consistent with \ab\ calculations, which report a large concentration of Berry curvature emanating along the Weyl loop \cite{Co3Sn2S2_Enke}. Further quantitative analysis is needed to better understand the relationship between the Weyl loop and the anomalous Hall and Nernst effects in \s\ as well as other topological magnets \cite{Co2MnGa_me, Co3Sn2S2_Shuang, Co3Sn2S2_ZeroFieldNernst}. In the future, it may also be exciting to explore other exotic responses of magnetic topological Weyl loops.

\ \\
The authors thank Diamond Light Source for access to Beamline I05 (SI17924, SI19313). T.A.C. acknowledges the support of the National Science Foundation Graduate Research Fellowship Program (DGE-1656466). T.N. and S.S.T. acknowledge support from the European Union Horizon 2020 Research and Innovation Program (ERC-StG-Neupert-757867-PARATOP). X.L. acknowledges financial support from the China Scholarship Council.

\clearpage

\clearpage

\clearpage


\begin{thebibliography}{10}

\bibitem{news_daSilvaNeto}
E.~H. da~Silva~Neto, ``{W}eyl''ing away time-reversal symmetry. {\it Science\/}
  {\bf 365}, 1248 (2019).

\bibitem{ReviewMagTopo_Tokura}
Y.~Tokura, K.~Yasuda, A.~Tsukazaki, {M}agnetic topological insulators. {\it
  Nat. Rev. Phys.\/} {\bf 1}, 126 (2019).

\bibitem{RMPWeylDirac_Armitage}
N.~P. Armitage, E.~J. Mele, A.~Vishwanath, {W}eyl and {D}irac semimetals in
  three-dimensional solids. {\it Rev. Mod. Phys.\/} {\bf 90}, 015001 (2018).

\bibitem{ReviewQuantumMaterials_Hsieh}
D.~N. Basov, R.~D. Averitt, D.~Hsieh, Towards properties on demand in quantum
  materials. {\it Nat. Mat.\/} {\bf 16}, 1077 (2017).

\bibitem{ReviewQuantumMaterials_Nagaosa}
Y.~Tokura, M.~Kawasaki, N.~Nagaosa, Emergent functions of quantum materials.
  {\it Nat. Phys.\/} {\bf 13}, 1056 (2017).

\bibitem{ARCMP_me}
M.~Z. Hasan, S.-Y. Xu, I.~Belopolski, S.-M. Huang, Discovery of {W}eyl fermion
  semimetals and topological {F}ermi arc states. {\it Ann. Rev. Cond. Matt.
  Phys.\/} {\bf 8}, 289 (2017).

\bibitem{ARCMP_Xue}
K.~He, Y.~Wang, Q.-K. Xue, {T}opological {M}aterials: {Q}uantum {A}nomalous
  {H}all {S}ystem. {\it Ann. Rev. Cond. Matt. Phys.\/} {\bf 9}, 329 (2018).

\bibitem{Fe3Sn2_Checkelsky}
L.~Ye, {\it et~al.\/}, Massive {D}irac fermions in a ferromagnetic kagome
  metal. {\it Nature\/} {\bf 555}, 638 (2018).

\bibitem{Fe3Sn2_Jiaxin}
J.-X. Yin, {\it et~al.\/}, Giant and anisotropic many-body spin-orbit
  tunability in a strongly correlated kagome magnet. {\it Nature\/}  (2018).

\bibitem{Co2MnGa_me}
I.~Belopolski, {\it et~al.\/}, Discovery of topological {W}eyl fermion lines
  and drumhead surface states in a room temperature magnet. {\it Science\/}
  {\bf 365}, 1278 (2019).

\bibitem{WeylLoopSuperconductor_Nandkishore}
Y.~Wang, R.~Nandkishore, Topological surface superconductivity in doped {W}eyl
  loop materials. {\it Phys. Rev. B\/} {\bf 95}, 060506 (2017).

\bibitem{WeylLines_Kane}
O.~Stenull, C.~L. Kane, T.~C. Lubensky, Topological phonons and {W}eyl lines in
  three dimensions. {\it Phys. Rev. Lett.\/} {\bf 117}, 068001 (2016).

\bibitem{WeylDiracLoop_Nandkishore}
R.~Nandkishore, {W}eyl and {D}irac loop superconductors. {\it Phys. Rev. B\/}
  {\bf 93}, 020506 (2016).

\bibitem{Co3Sn2S2_HechangLei}
Q.~Wang, {\it et~al.\/}, Large intrinsic anomalous {H}all effect in
  half-metallic ferromagnet {Co$_3$Sn$_2$S$_2$} with magnetic {W}eyl fermions.
  {\it Nat. Commun.\/} {\bf 9}, 3681 (2018).

\bibitem{Co3Sn2S2_Enke}
E.~Liu, {\it et~al.\/}, Giant anomalous {H}all effect in a ferromagnetic
  kagome-lattice semimetal. {\it Nat. Phys.\/} {\bf 14}, 1125 (2018).

\bibitem{Co3Sn2S2_Zurab}
Z.~Guguchia, {\it et~al.\/}, Tunable anomalous {H}all conductivity through
  volume-wise magnetic competition in a topological kagome magnet. {\it Nat.
  Commun.\/} {\bf 11}, 559 (2020).

\bibitem{Co3Sn2S2_Sobany}
P.~Vaqueiro, G.~G. Sobany, A powder neutron diffraction study of the metallic
  ferromagnet {C}o$_3${S}n$_2${S}$_2$. {\it Solid State Sci.\/} {\bf 11}, 513
  (2009).

\bibitem{Co3Sn2S2_Weihrich}
W.~Schnelle, {\it et~al.\/}, Ferromagnetic ordering and half-metallic state of
  {S}n$_2${C}o$_3${S}$_2$ with the shandite-type structure. {\it Phys. Rev.
  B\/} {\bf 88}, 144404 (2013).

\bibitem{Co3Sn2S2_QiunanXu}
Q.~Xu, {\it et~al.\/}, Topological surface {F}ermi arcs in the magnetic {W}eyl
  semimetal {Co$_3$Sn$_2$S$_2$}. {\it Phys. Rev. B\/} {\bf 97}, 235416 (2018).

\bibitem{Co3Sn2S2_Jiaxin}
J.-X. Yin, {\it et~al.\/}, Negative flat band magnetism in a spin-orbit-coupled
  correlated kagome magnet. {\it Nat. Phys.\/} {\bf 15}, 443 (2019).

\bibitem{Co3Sn2S2_YulinChen}
D.~F. Liu, {\it et~al.\/}, Magnetic {W}eyl semimetal phase in a {K}agom\'e
  crystal. {\it Science\/} {\bf 365}, 1282 (2019).

\bibitem{Co3Sn2S2_Beidenkopf}
N.~Morali, {\it et~al.\/}, Fermi-arc diversity on surface terminations of the
  magnetic {W}eyl semimetal {C}o$_3${S}n$_2${S}$_2$. {\it Science\/} {\bf 365},
  1286 (2019).

\bibitem{Co3Sn2S2_me_APS2019}
I.~Belopolski, {\it et~al.\/}, Investigation of the magnetic {Weyl} semimetal
  candidate {C}o$_3${S}n$_2${S}$_2$ by {ARPES}. {\it American Physical Society
  March Meeting\/} (2019). E01.00004.

\bibitem{Diamond_Hoesch}
M.~Hoesch, {\it et~al.\/}, A facility for the analysis of the electronic
  structures of solids and their surfaces by synchrotron radiation
  photoelectron spectroscopy. {\it Rev. Sci. Inst.\/} {\bf 88}, 013106 (2017).

\bibitem{DFT2}
G.~Kresse, J.~Furthmueller, Efficient iterative schemes for \textit{ab initio}
  total-energy calculations using a plane-wave basis set. {\it Phys. Rev. B\/}
  {\bf 54}, 11169 (1996).

\bibitem{DFT_Efficiency_KressFurthmueller}
G.~Kresse, J.~Furthmueller, Efficiency of ab-initio total energy calculations
  for metals and semiconductors using a plane-wave basis set. {\it Comp. Mat.
  Sci.\/} {\bf 6}, 15 (1996).

\bibitem{DFT4}
J.~P. Perdew, K.~Burke, M.~Ernzerhof, Generalized gradient approximation made
  simple. {\it Phys. Rev. Lett.\/} {\bf 77}, 3865 (1996).

\bibitem{Yates_Wannier90}
G.~Pizzi, {\it et~al.\/}, Wannier90 as a community code: new features and
  applications. {\it J. Phys.: Cond. Mat.\/} {\bf 32}, 165902 (2020).

\bibitem{Soluyanov_WannierTools}
Q.~S. Wu, S.~Zhang, H.-F. Song, M.~Troyer, A.~A. Soluyanov, {W}annier{T}ools:
  an open-source software package for novel topological materials. {\it Comput.
  Phys. Commun.\/} {\bf 224}, 405 (2018).

\bibitem{NbP_me}
I.~Belopolski, {\it et~al.\/}, Criteria for directly detecting topological
  {F}ermi arcs in {W}eyl semimetals. {\it Phys. Rev. Lett.\/} {\bf 116}, 066802
  (2016).

\bibitem{HgTe_ChangLiu}
C.~Liu, {\it et~al.\/}, Tunable spin helical {D}irac quasiparticles on the
  surface of three-dimensional {H}g{T}e. {\it Phys. Rev. B\/} {\bf 92}, 115436
  (2015).

\bibitem{Co3Sn2S2_Shuang}
H.~Zhou, {E}nhanced anomalous {H}all effect in the magnetic topological
  semimetal {C}o$_3${S}n$_{2-x}${I}n$_x${S}$_2$. {\it Phys. Rev. B\/} {\bf
  101}, 125121 (2020).

\bibitem{Co3Sn2S2_ZeroFieldNernst}
S.~N. Guin, {\it et~al.\/}, {Z}ero-{F}ield {N}ernst {E}ffect in a
  {F}erromagnetic {K}agome-{L}attice {W}eyl-{S}emimetal
  {C}o$_3${S}n$_2${S}$_2$. {\it Adv. Mat.\/} {\bf 31}, 1806622 (2019).

\end{thebibliography}
\end{document}